\documentclass[conference]{IEEEtran}
\IEEEoverridecommandlockouts
\usepackage{cite}
\usepackage{amsmath,amssymb,amsfonts}
\usepackage{graphicx}
\usepackage{lipsum}
\usepackage{textcomp}
\usepackage{algorithm}
\usepackage{tikz}
\usetikzlibrary{patterns}

\usepackage{array}

\usepackage{booktabs} 
\usepackage{algpseudocode}
\usepackage{xcolor}
\def\BibTeX{{\rm B\kern-.05em{\sc i\kern-.025em b}\kern-.08em
    T\kern-.1667em\lower.7ex\hbox{E}\kern-.125emX}}
\usepackage{comment}
\usepackage{makecell}
\usepackage{amsmath}
\usepackage{amsmath, amssymb, amscd, amsthm, amsfonts}
\usepackage{colortbl}
\usepackage{soul}
\usepackage{multicol}

\usepackage[colorlinks=true,linkcolor=blue,anchorcolor=black,citecolor=red,filecolor=black,menucolor=black,runcolor=black,urlcolor=black]{hyperref}
    
\begin{document}



\title{
A Novel Generative AI-Based Framework for Anomaly Detection in Multicast Messages in Smart Grid Communications
\\



\thanks{A. Zaboli and J. Hong are with the Department of Electrical and Computer Engineering, University of Michigan -- Dearborn, Dearborn, MI 48128, USA.

S. L. Choi is with the Power Systems Engineering Center, National Renewable Energy Laboratory (NREL), Golden, CO 80401, USA.

T.-J. Song is with the Department of Urban Engineering, Chungbuk National University, Cheongju 28644, South Korea.

}
}

\author{\IEEEauthorblockN{Aydin Zaboli, \textit{Student Member, IEEE}, Seong Lok Choi, \textit{Member, IEEE}, Tai-Jin Song, \textit{Member, IEEE}, \\ Junho Hong, \textit{Senior Member, IEEE}
}}

\maketitle

\begin{abstract}
Cybersecurity breaches in digital substations can pose significant challenges to the stability and reliability of power system operations. To address these challenges, defense and mitigation techniques are required. Identifying and detecting anomalies in information and communication technology (ICT) is crucial to ensure secure device interactions within digital substations. This paper proposes a task-oriented dialogue (ToD) system for anomaly detection (AD) in datasets of multicast messages e.g., generic object oriented substation event (GOOSE) and sampled value (SV) in digital substations using large language models (LLMs). 
This model has a lower potential error and better scalability and adaptability than a process that considers the cybersecurity guidelines recommended by humans, known as the human-in-the-loop (HITL) process.
Also, this methodology significantly reduces the effort required when addressing new cyber threats or anomalies compared with machine learning (ML) techniques, since it leaves the model's complexity and precision unaffected and offers a faster implementation. These findings present a comparative assessment, conducted utilizing standard and advanced performance evaluation metrics for the proposed AD framework and the HITL process. To generate and extract datasets of IEC 61850 communications, a hardware-in-the-loop (HIL) testbed was employed.
\end{abstract}


\section{Introduction} \label{Introduction}
IEC 61850-based digital substations occupy an essential position within the power grid framework by managing the distribution, transformation, and combination of energy flows. The emergence of smart grids has led to the fusion of the power grid infrastructure with communication networks and computational capabilities, introducing a wide array of innovative applications such as automated data collection and the remote control of electrical systems and components~\cite{sun2018cyber}. Nonetheless, this combination brings a variety of security vulnerabilities to the smart grid. Intrusion detection systems (IDSs) are instrumental in detecting malicious activities and curtailing the actions of adversaries. Notably, IDSs have found extensive applications in conventional ICT domains to fulfill this objective. However, adopting IEC 61850 and introducing communication protocols, including GOOSE and SV (i.e., multicast messages), have paved the way for specialized malicious strategies that manifest distinct traffic and attack configurations. These could encompass replay attacks, unauthorized data injection, and denial of service (DoS) attacks. Consequently, there is a pressing need for IDSs to acquire new signatures for purposes of training, testing, validating, and evaluating their efficacy in the face of challenges~\cite{10339874, hong2022automated}.

ML techniques employed in IDSs have become crucial in detecting and addressing anomalies within GOOSE and SV multicast messages. These methodologies are noted for their precision and data-centric approach, offering a sophisticated framework for cybersecurity measures, but they still have challenges. A principal drawback is the necessity for continuous model re-training in response to newly emerging attack vectors. Each time an innovative pattern of attack is identified, an update of the ML models is required to encompass this new information. This re-training process is time-consuming and resource-intensive, creating a temporal window of vulnerability wherein the system remains susceptible to new threats not yet incorporated into the model’s intelligence base~\cite{beg2023review}. Moreover, the scalability of these IDS-based ML models, their decision-making efficacy, and the efficiency of data processing mechanisms assume critical importance in the operational dynamics of AD. The scalability issue pertains to the model's ability to adapt and maintain performance levels as the network size or data volume expands. The decision-making process involves the model's capability to distinguish accurately between secure and malicious activities, a task that becomes increasingly complex with the evolution of sophisticated attack techniques. Lastly, the data processing aspect highlights the need for efficient handling and analysis of vast datasets that systems often rely upon~\cite{hong2017intelligent, chen2016modeling}. 

Considering these points, developing more adaptive, resilient, and scalable AD solutions is imperative. These solutions should minimize the latency in incorporating new threat intelligence and enhance the decision-making processes and data-handling capacities to better manage the complexities associated with GOOSE and SV communications. 
Hence, LLMs (e.g., Anthropic Claude Pro~\cite{anthropic}, Microsoft Copilot AI~\cite{Copilot}) can present a more flexible methodology compared with traditional ML frameworks and an HITL process. Distinctively engineered to understand the details of context, LLMs possess the inherent capability to detect novel attacks/anomalies without the prerequisite of prior training in these scenarios. This adeptness at contextual comprehension significantly reduces the effort and resources typically necessitated by the continual evolution of attacks. Rather than undergoing repetitive cycles of re-training, LLMs are capable of interpreting and adjusting to new data autonomously, thereby offering a more robust and efficient AD method within the context of digital substations~\cite{gill2023chatgpt, ten2011anomaly}. This innovative approach emphasizes the potential of LLMs to enhance cybersecurity by providing a tool that can dynamically evolve in response to cyberattacks.

\subsection{Related Work}
A progressive shift toward leveraging sophisticated algorithms has revealed a desire to enhance the AD process to secure digital substations based on ML models. These models can enhance intrusion detection in substations by analyzing patterns and anomalies in datasets. This allows for the real-time detection of cyberattacks, ensuring grid stability and security~\cite{reda2021vulnerability, zhu2020intrusion, choi2020multi, kreimel2020anomaly, wang2022anomaly}. Alvee \textit{et al.}~\cite{alvee2021ransomware} proposed an AI-based ransomware detection approach that utilizes a convolutional neural network (CNN). The unique aspect of this approach is the conversion of binary files into 2-D image files for detection. Experimental results indicate a high detection accuracy of $96.22\%$. However, their dataset is not based on the HIL testbed, and different attack scenarios are neglected which makes the efficiency of the approach controversial. An advanced ML technique for real-time AD was proposed in~\cite{panthi2020anomaly}. It employed feature extraction techniques to optimize the data; however, handling the vast amount of data without compromising the system's performance is challenging. Moreover, a trade-off between the detection accuracy, computational efficiency, and robustness against evolving cyberattack patterns is questionable. A combination of CNN and long short-term memory (LSTM) networks was proposed for enhanced performance in~\cite{ankitdeshpandey2020development}. Investigating techniques to improve the robustness and generalization of models to new attack types and evolving cyber threats can be a challenge. Also, feature selection and engineering to identify the most critical features for an AD is arguable. Upadhyay \textit{et al.}~\cite{upadhyay2020gradient} introduced a gradient boosting method for the feature selection in an IDS that demonstrates improved accuracy and reduced false positives (FPs) in detecting anomalies. Nonetheless, they considered a general attack vector for the performance analysis process; hence, there is not sufficient information to distinguish the different types of attacks. A novel IDS was designed to combat manufacturing message specification (MMS)-based measurement attacks that integrated advanced detection algorithms to improve the accuracy~\cite{zhu2020intrusion}. Ustun \textit{et al.}~\cite{ustun2021machine} suggested a novel IDS that employed a combination of CNN and LSTM networks to capture spatial and temporal patterns in data. Future research could also look into the scalability of proposed IDS solutions, ensuring that they can be effectively integrated into various smart grid environments without significant modifications.

According to literature surveys, there is no research to consider a directed method based on human recommendations known as ``CyberGridToD'' (\textbf{Cyber}security of Smart \textbf{Grid}s using a \textbf{T}ask-\textbf{O}riented \textbf{D}ialogue system), for attacks/anomalies using LLMs. Most research has issues of scalability, adaptability, robustness, and time processing. Hence, a framework that can handle challenges with less effort, without the involvement of a human expert at each step, is needed.
\subsection{Contributions}
This paper proposes an extended ToD system called CyberGridToD, designed based on LLMs (e.g., Anthropic Claude Pro), that has advantages over HITL processes using LLMs in the area of generative artificial intelligence (GenAI) and ML algorithms. It is built on historical human recommendation data, enabling it to automate decision-making processes by emulating human decision patterns, potentially leading to a lower error rate over time as it learns from new data. The learning capabilities allow it to be continuously improved. This system can respond quickly due to its automated nature, although this can vary with complexities. It excels in data processing, being capable of handling and analyzing large volumes of data. CyberGridToD is highly scalable and able to serve many users simultaneously without performance degradation. It requires time to build user trust in its recommendations and may struggle with highly complex or situations not covered in its training data. Thus, the main strengths of the proposed framework can be summarized as follows:
\begin{itemize}
    \item It pioneered for the first time, the application of an LLM-based ToD system for efficient and reliable AD in multicast messages within digital substations, offering a novel approach to enhancing the security and stability of smart grid operations.
    \item A rigorous evaluation of the proposed framework was conducted using standard and advanced metrics, addressing the limitations of previous research and establishing a new benchmark for assessing the performance of IDSs with less effort and more adaptability and scalability.
\end{itemize}
\subsection{Paper Outline}
The rest of this paper is organized as follows: Section~\ref{SectionII} overviews the IEC 61850-based GOOSE and SV protocols and datasets as well as human recommendations for multicast messages. The proposed framework for AD in the GOOSE and SV datasets is represented in Section~\ref{Proposed}. Section~\ref{Results} presents the results and discussion with a comparison between the HITL  process and the proposed framework. Finally, conclusions and directions for future work are outlined in Section~\ref{Conclusion}.
\section{Cybersecurity in Digital Substations} \label{SectionII}
As digital substations integrate advanced communication technologies, they become potential targets for cyberattacks. Ensuring the cybersecurity of critical infrastructure involves implementing multi-layered protective measures, ranging from physical access controls to advanced IDSs. Regular assessments are essential to tackle evolving cyberattacks and attain reliability and stability in power grids~\cite{hong2014integrated, roomi2023analysis}. 
A cyber-physical power system testbed is an essential infrastructure for examining in a controlled realistic setting the causal relationships between cyber intrusions and the resilience and reliability of power systems. The real-time HIL testbed integrates various components, including hardware, software, communication protocols, and simulation technologies, all synchronized with the global positioning system (GPS). Such integration is crucial for investigating the real-time dynamics of communications and information processing, which are vital for analyzing cyber intrusions, enhancing detection capabilities, and formulating effective mitigation strategies~\cite{hong2021implementation}.

The architecture of the HIL testbed encompasses a wide range of components, including protective intelligent electronic devices (IEDs), software-defined networking (SDN) switches, GPS, merging unit IED, a supervisory control and data acquisition (SCADA) system, a real-time digital simulator, and an amplifier, details of which will be elaborated. The system employs a distributed management system (DMS) SCADA framework for acquiring measurements and executing control commands through DNP3 protocols. Implemented IEDs have the capability to send control signals to circuit breakers (CBs), while a CB (with actuator) is designed to respond to GOOSE messages by transmitting its status (i.e., open or closed) back to the protective IEDs. Additionally, the merging unit IED is tasked with relaying digital current and voltage readings from the digital real-time simulator to the protective IEDs, leveraging the amplifier~\cite{ten2011anomaly, choi2020multi}. The proposed LLM-based framework is designed to detect anomalies and security threats within the multicast messages, maintaining connectivity with SDN switches to ensure comprehensive protection. The upcoming sections represent the GOOSE and SV datasets from the HIL testbed and their feature extraction process, along with defined anomaly recommendations included in the proposed LLM-based ToD framework.
\subsection{IEC 61850-based Multicast Messages Dataset}
\begin{figure}[!t]
\centerline{\includegraphics[width=0.8\columnwidth]{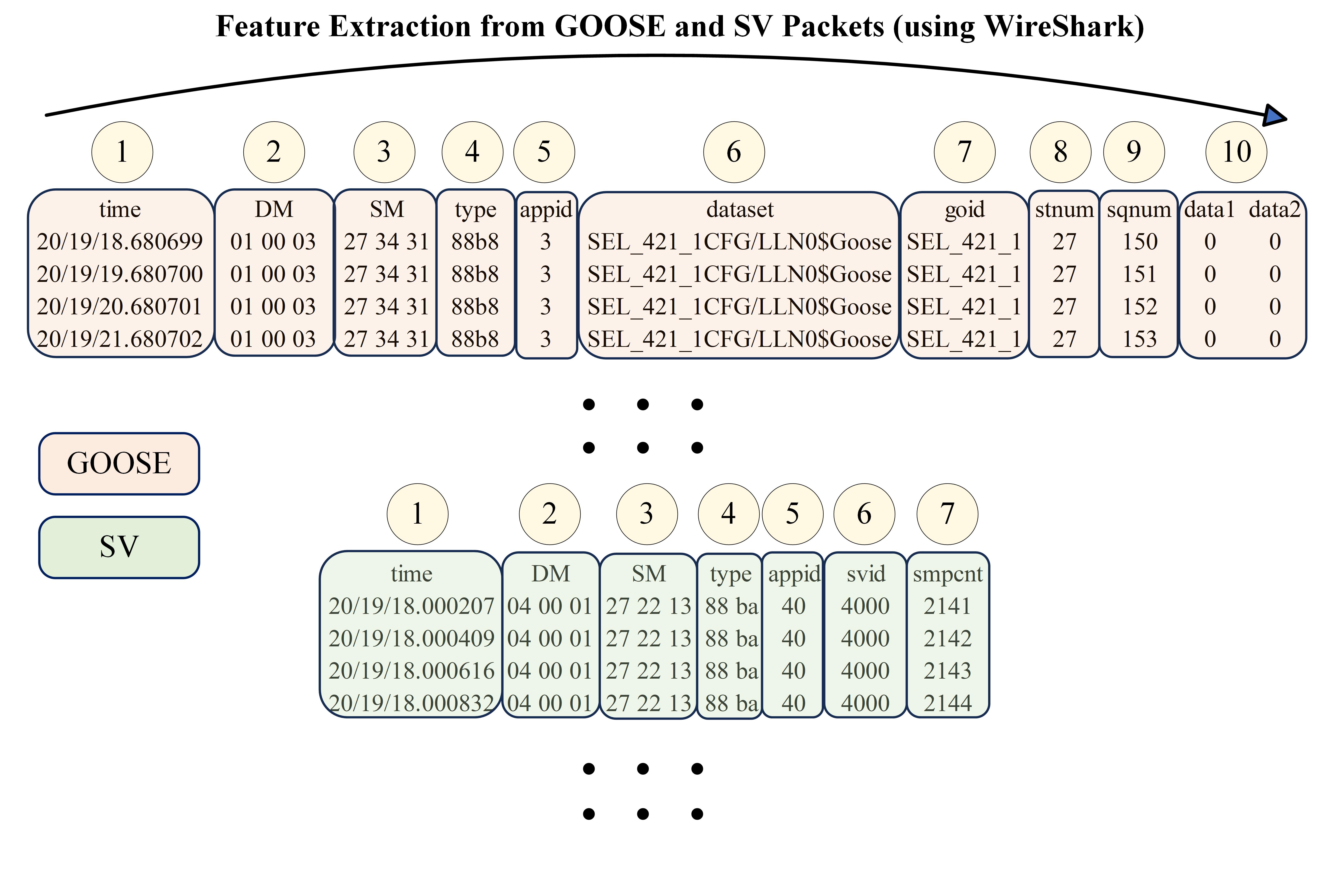}}
\caption{A pre-processing step based on the feature extraction for a log of GOOSE and SV messages (actual data from an HIL testbed).}
\label{fig:Pre-processing}
\end{figure}

In the HIL testbed, the extraction of GOOSE and SV packets is meticulously performed. Utilizing Wireshark, a network packet analyzer, the capture of packets is facilitated, as illustrated in the Fig.~\ref{fig:Pre-processing}. This process involves active monitoring and analysis of network traffic within the HIL testbed, allowing for the detailed observation and documentation of the packet flow. Through Wireshark's capabilities, researchers are able to obtain a comprehensive snapshot of the communication patterns, thereby enabling a better understanding of interactions between different components within the HIL testbed. This methodological approach ensures the accurate extraction of critical packets and enhances the overall research process by providing valuable insights into the operational dynamics of cyber-physical systems.
Analysis reveals that the GOOSE packet data encompasses ten distinct data types, designated by the features extracted. Similarly, the SV dataset adheres to similar methodologies, emphasizing seven critical features identified as dataset columns. The \textit{Time} attribute records the precise moment of packet transmission, formatted to include hours, minutes, seconds, and even microseconds for in-depth accuracy. The terms \textit{DM} and \textit{SM} are acronyms for destination and source media access control (MAC) addresses, respectively, serving as critical identifiers within the communication process. Specifically, the \textit{DM} address for GOOSE messages, denoted as ($01 00 03$), targets the devices subscribed to this MAC address, while the \textit{SM} address, represented as $27 34 31$, identifies the sending IED in this example. The classification of GOOSE and SV packets is further refined by the \textit{type} indicator, assigned values of $88 b8$ and $88 ba$, respectively. Additionally, the \textit{APPID} values for GOOSE and SV communications are designated as $3$ and $40$, sequentially. The \textit{dataset} and \textit{goID} attributes, contingent upon the \textit{DM} address, specify the dataset name and GOOSE identification. Also, \textit{stNum} and \textit{sqNum} represent the state and sequence numbers within GOOSE communications. Moreover, the analysis incorporates two data types, \textit{data1} and \textit{data2}, extracted from features of GOOSE packets. Please note that the number of data and data types will vary based on the substation design and engineering process. Within the SV dataset, the \textit{savPdu} field contains \textit{svID} and \textit{smpCnt}, denoting SV identification and the sample count number. The selection of 10 features in GOOSE and 7 features in SV is due to their more significant impact and importance in substation operation compared with other data types in GOOSE and SV. The next section presents potential anomaly recommendations for these datasets according to their features.
\subsection{Anomaly Recommendations in Digital Substations}
\begin{figure}[!t]
\centerline{\includegraphics[width=1.0\columnwidth]{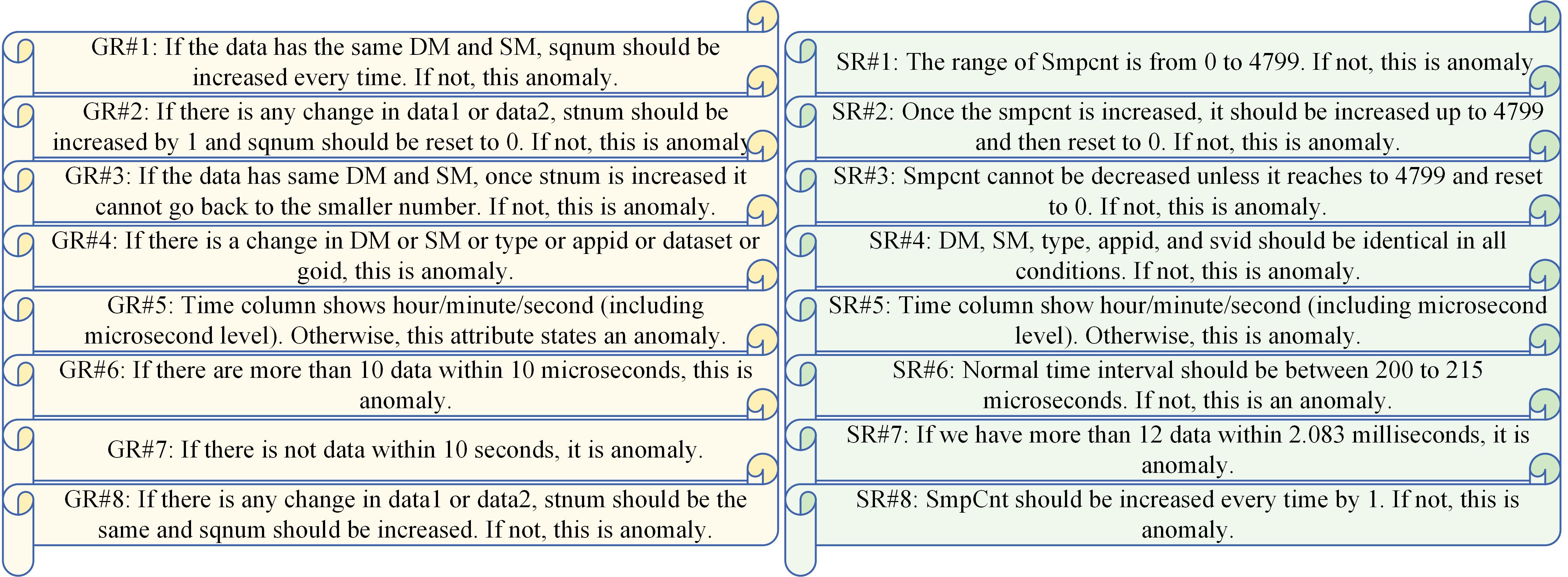}}
\caption{GOOSE/SV recommendations in the HITL and proposed frameworks.}
\label{fig:recommendations}
\end{figure}

This section presents all anomaly considerations for GOOSE and SV messages, including $8$ recommendations each, as shown in Fig.~\ref{fig:recommendations}. 
As can be seen, GR\# and SR\# demonstrate the GOOSE and SV anomaly recommendations, respectively, according to their datasets~\cite{hong2014integrated}. The details of AD processes for GOOSE and SV datasets are illustrated in Algorithms~\ref{GOOSErecom} and \ref{SVrecom}, separately.
\begin{algorithm}[!h]
\scriptsize
\caption{Anomaly Detection in GOOSE Dataset.} \label{GOOSErecom}
\textbf{Input:} Set of GOOSE packets $G$, Recommendations $GR$

\textbf{Output:} Set of anomalies $A_{G}$

\begin{algorithmic}[]
\State Initialize $A_{G} \leftarrow \emptyset$
\For{$i = 2$ \textbf{to} $|G|$}
    \State $p \leftarrow G[i]$, $p_{prev} \leftarrow G[i-1]$
    \If{$DM(p) = DM(p_{prev}) \land SM(p) = SM(p_{prev})$}
        \If{$sqnum(p) \neq sqnum(p_{prev}) + 1$}
            \State $A_{G} \leftarrow A \cup \{\text{``sqnum anomaly''}\}$
        \EndIf
        \If{$stnum(p) < stnum(p_{prev})$}
            \State $A_{G} \leftarrow A_{G} \cup \{\text{``stnum decrease anomaly''}\}$
        \EndIf
    \EndIf
    \If{$data(p) \neq data(p_{prev})$} 
        \If{$stnum(p) \neq stnum(p_{prev}) + 1 \lor sqnum(p) \neq 0$}
            \State $A_{G} \leftarrow A_{G} \cup \{\text{``stnum/sqnum reset anomaly''}\}$
        \EndIf
    \EndIf
    \If{$\exists attr \in \{DM, SM, type, appid, dataset, goid\}: attr(p) \neq attr(p_{prev})$}
        \State $A_{G} \leftarrow A_{G} \cup \{\text{``attribute change anomaly''}\}$
    \EndIf
\EndFor
\State $T \leftarrow \text{time column of each packet in } G$
\If{$\neg \text{isValidFormat}(T)$}
    \State $A_{G} \leftarrow A_{G} \cup \{\text{``time format anomaly''}\}$
\EndIf
\If{$\exists$ sequence in $G$ with $>10$ packets within $10\mu s$}
    \State $A_{G} \leftarrow A_{G} \cup \{\text{``high data rate anomaly''}\}$
\EndIf
\If{$\exists$ gap $>10s$ without data between consecutive packets in $G$}
    \State $A_{G} \leftarrow A_{G} \cup \{\text{``data gap anomaly''}\}$
\EndIf
\For{$i = 2$ \textbf{to} $|G|$}
    \If{$data(G[i]) \neq data(G[i-1])$ \textbf{and} $(stnum(G[i]) \neq stnum(G[i-1]) \lor sqnum(G[i]) \neq sqnum(G[i-1]) + 1)$}
        \State $A_{G} \leftarrow A_{G} \cup \{\text{``data change anomaly''}\}$
    \EndIf
\EndFor
\State \Return $A_{G}$
\end{algorithmic}
\end{algorithm}
The algorithm inputs a set of GOOSE packets $G$ and a set of recommendations $GR$ and outputs a set of identified anomalies $A_{G}$, according to Algorithm~\ref{GOOSErecom}. This algorithm begins by initializing an empty set $A_{G}$ to store anomalies. It iterates through each GOOSE packet $p$ in the set $G$, starting from the second packet (index 2), comparing each packet $p$ with its predecessor $p_{prev}$. In terms of checking anomalies, if the destination MAC address (DM) and source MAC address (SM) of $p$ and $p_{prev}$ are identical but the sequence number $sqnum$ of $p$ is not exactly one more than $sqnum$ of $p_{prev}$, an ``$sqnum$ anomaly'' is added to $A_{G}$. If $stnum$ of $p$ is less than $stnum$ of $p_{prev}$, an ``$stnum$ decrease anomaly'' is added to $A$. If there is a change in data between $p$ and $p_{prev}$, and either the status number $stnum$ is not incremented or the $sqnum$ is not reset to $0$, an ``$stnum$/$sqnum$ reset anomaly'' is added to $A$. If any attribute among DM, SM, type, appid, dataset, or goid changes from $p_{prev}$ to $p$, an ``attribute change anomaly'' is noted in $A_{G}$. Regarding the time format, the time column $T$ of each packet in $G$ is checked for correct formatting. If any packet's time format is incorrect, a ``time format anomaly'' is added to $A_{G}$. The algorithm checks for more than $10$ packets occurring within 10 $\mu s$ and, if found, adds a ``high data rate anomaly'' to $A_{G}$. It checks for a gap of more than 10 $s$ without data between two consecutive packets, adding a ``data gap anomaly'' to $A_{G}$ if such a gap is found. Please note that these time settings can be adjusted for different IEC 61850 configurations. In a separate loop, the algorithm examines each packet for changes in data between consecutive packets. If there is a change without the corresponding increment in $stnum$ or $sqnum$, a ``data change without $sqnum$ increment anomaly'' is added to $A_{G}$. Finally, the set $A_{G}$ containing all anomalies is returned in the output.
\begin{algorithm}[!h]
\scriptsize
\caption{Anomaly Detection in SV Dataset.} \label{SVrecom}
\textbf{Input:} Set of SV packets $SV$, Recommendations $SVR$

\textbf{Output:} Set of anomalies $A_{SV}$
\begin{algorithmic}[]
\State Let \( SmpCnt \in \{0, 1, \ldots, 4799\} \)
\State \( SmpCnt_{prev} \gets -1 \)
\State \( {A_{SV}} \gets \emptyset \) \Comment{Initialize the set of anomalies}
\For{\( packet \) in \( \text{SV packets} \)}
    \If{\( packet.SmpCnt < 0 \) \textbf{or} \( packet.SmpCnt > 4799 \)}
        \State \( {A_{SV}} \gets {A_{SV}} \cup \{\text{``SmpCnt range anomaly''}\} \)
    \EndIf
    \If{\( SmpCnt_{prev} \geq 0 \)}
        \If{\( SmpCnt_{prev} \neq 4799 \) \textbf{and} \( packet.SmpCnt \neq SmpCnt_{prev} + 1 \)}
            \State \( {A_{SV}} \gets {A_{SV}} \cup \{\text{``SmpCnt increase anomaly''}\} \)
        \EndIf
        \If{\( SmpCnt_{prev} < 4799 \) \textbf{and} \( packet.SmpCnt < SmpCnt_{prev} \)}
            \State \( {A_{SV}} \gets {A_{SV}} \cup \{\text{``SmpCnt decrease anomaly''}\} \)
        \EndIf
    \EndIf
    \If{\( \exists \) field \( f \in \{DM, SM, Type, AppID, SVID\} \) such that \( f_{packet} \neq f_{prev} \)}
        \State \( {A_{SV}} \gets {A_{SV}} \cup \{\text{``Field consistency anomaly''}\} \)
    \EndIf
    \If{\( \text{Format}(packet.Time) \neq \text{``HH:MM:SS.ssssss''} \)}
        \State \( {A_{SV}} \gets {A_{SV}} \cup \{\text{``Time format anomaly''}\} \)
    \EndIf
    \If{\( \text{Interval}(packet.Time) \notin [200\mu s, 215\mu s] \)}
        \State \( {A_{SV}} \gets {A_{SV}} \cup \{\text{``Time interval anomaly''}\} \)
    \EndIf
    \If{\( \text{Count}( \text{Data within } 2.083 ms) > 12 \)}
        \State \( {A_{SV}} \gets {A_{SV}} \cup \{\text{``Data rate anomaly''}\} \)
    \EndIf
    \State \( SmpCnt_{prev} \gets packet.SmpCnt \)
\EndFor
\If{\( {A_{SV}} \neq \emptyset \)}
    \State Report \( {A_{SV}} \)
\EndIf
\end{algorithmic}
\end{algorithm}

Algorithm~\ref{SVrecom} states an AD process for SV datasets according to different recommendations. Inputs and outputs are defined clearly, and sample count $smpcnt$ is defined to be within the inclusive range of $0$ to $4799$. This is the valid range of the $smpcnt$ value. $smpcnt_{prev}$ is initialized to $-1$, indicating that at the start, there is no previous sample count to compare with. A set $A_{SV}$ is initialized to be empty. This set will hold any detected anomalies. The algorithm iterates over each $packet$ within the collection of GOOSE packets. If $packet.SmpCnt$ is less than $0$ or greater than $4799$, an anomaly string ``SmpCnt range anomaly'' is added to the set $A_{SV}$, indicating that the sample count is out of the valid range. If $SmpCnt_{prev}$ is not $-1$ (which means that this is not the first packet), and if $SmpCnt_{prev}$ is not $4799$ and 
$packet.SmpCnt$ is not equal to $SmpCnt_{prev}+1$, an ``SmpCnt increase anomaly'' is added to $A_{SV}$. This checks for the correct sequential increase of $SmpCnt$. If $SmpCnt_{prev}$ is less than $4799$ and $packet.SmpCnt$ is less than $SmpCnt_{prev}$, an ``SmpCnt decrease anomaly'' is added to $A_{SV}$. This checks that $SmpCnt$ does not decrease before reaching the maximum value and resetting. The algorithm checks for consistency in fields $DM$, $SM$, $Type$, $AppID$, and $SVID$. If there exists a field $f$ such that the current packet's field $f_{packet}$ is not equal to the previous packet's field $f_{prev}$, a ``field consistency anomaly'' is added to $A_{SV}$. The algorithm checks if the format of $packet.Time$ matches the expected format, ``HH:MM:SS.ssssss.'' If not, a ``time format anomaly'' is added to $A_{SV}$. The next step checks if the time interval of $packet.Time$ is not within the range of $200$ to $215$ $\mu s$. If it is outside this range, a ``time interval anomaly'' is added to $A_{SV}$. It then checks if there are more than $12$ packets within a $2.083$ $ms$ window. If so, a ``data rate anomaly'' is added to $A_{SV}$. After checking the current packet, $SmpCnt_{prev}$ is updated to the current packet's $SmpCnt$ for comparison with the next packet. Finally, if the set $A_{SV}$ is not empty, which means one or more anomalies have been detected, the set $A_{SV}$ is reported.
The next section will present the proposed framework using the SV and GOOSE datasets and defined human recommendations.

\section{A Proposed Large Language Model-based Task-Oriented Dialogue IDS Framework} \label{Proposed}
\begin{figure}[!t]
\centerline{\includegraphics[width=0.8\columnwidth]{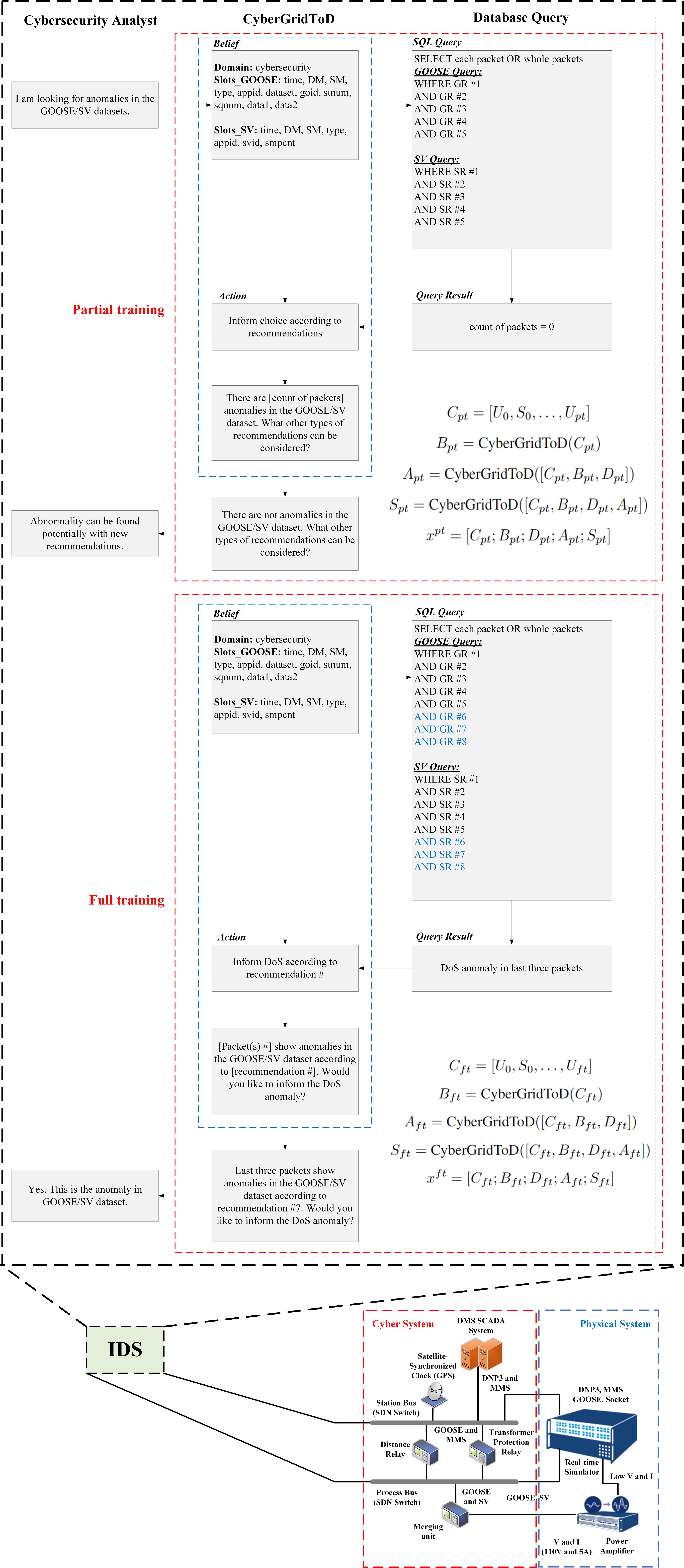}}
\caption{A proposed LLM-based ToD system for AD of multicast messages.}
\label{fig:ToDframework}
\end{figure}

ToD systems are typically assessed based on their proficiency in distinct subtasks, including tracking the dialogue state (known as belief state), managing the dialogue (encompassing action and decision prediction), and generating appropriate responses using an SQL query database. This subdivision of tasks has facilitated the development of specialized models for each sub-task, a methodology that has become prevalent in the field. This research investigates the efficacy of a unified, end-to-end model known as ``CyberGridToD'' for managing these tasks, as illustrated in Fig.~\ref{fig:ToDframework}.
The cybersecurity analyst component is the LLM (e.g., Anthropic Claude Pro), which processes the GOOSE and SV data and information to detect anomalies. It takes the packets, CyberGridToD labels, and anomaly scores as inputs to learn the characteristics of anomalies. Then, it utilizes this learned knowledge to analyze new GOOSE and SV data and identify potential anomalies.

In the context of dialogues, each interaction is composed of successive turns. At a given turn ``$t$,'' the user inputs a statement (denoted as $U_{t}$), to which the system responds (denoted as $S_{t}$). When generating a response, particularly during the inference phase, CyberGridToD incorporates all preceding dialogue turns as contextual information, represented as $C_{t} = [U_{0}; S_{0}; ...; U_{t}]$. Then, this model produces a belief state, $B_{t}$ at each turn. The belief state is a compilation of triplets that capture specific values for various slots within a given domain, structured in the format (domain, slot name, value). The domain can be defined as ``cybersecurity,'' and there are two different slots for GOOSE and SV multicast messages including their features (e.g., time, DM, SM, and appid) in the relevant datasets. $A_{t}$ shows the action based on the contextual information, belief state, and domain. Finally, the result, $x^{t}$, is a tuple of different parameters included in the proposed framework. This framework and the proposed human recommendations are defined in LLMs to detect anomalies in the SV and GOOSE datasets. The SQL query block contains different human recommendations defined for these messages. The human recommendations shown in Fig.~\ref{fig:recommendations} are included by dividing them into two parts: the partial training and full training levels. Also, SQL queries are not considered for cases without human recommendations. In this query, 5 and 8 recommendations are considered for the partial and full training levels, respectively. Then, the proper results can be presented based on the iterative loop of directed dialogues. 

\subsection{Challenges of HITL Process and ML Models} \label{challenges}
A comparative analysis of three general AD methods (i.e., HITL, ML models, and the LLM-based ToD method) reveals the superior performance and adaptability of the proposed LLM-based ToD approach. While HITL relies on human operator availability and expertise (which may be limited by operator capabilities and inconsistent decision-making processes) ML models (though faster) are constrained by predefined input/output formats and limited adaptability, requiring retraining on new data or edge cases. In contrast, the proposed LLM-based ToD model offers a more holistic and efficient solution, leveraging natural language dialogue to allow for complex queries and responses, high adaptability to new scenarios and anomalies through the prompt engineering process, and the ability to provide explainable and transparent insights~\cite{chung2023instructtods, hu2024dialight}. More comprehensive detail along with a comparative description of these method is presented in Table~\ref{model_comparison}.
\begin{table*}[htbp]
\centering
\tiny
\caption{A comparison of LLM-based ToD, HITL, and ML models for GOOSE/SV dataset AD~\cite{chung2023instructtods, hu2024dialight, li2024personal, al2021review, nassif2021machine}.}
\label{model_comparison}
\begin{tabular}{|c|p{4.5cm}|p{4.5cm}|p{5cm}|}
\hline
\makecell{\textbf{Criteria}} & \makecell{\textbf{HITL}} & \makecell{\textbf{ML Models}} & \makecell{\textbf{Proposed (LLM-based ToD)}}  \\
\hline
\makecell{Interaction Mode} & Human-driven interaction, may be limited by operator availability and expertise. & Predefined input/output formats, limited interaction capabilities. & Natural language dialogue, allowing for complex queries and responses. \\
\hline
\makecell{Response Time} & Depends on human operator availability and speed, may have delays. & Generally fast, but may slow down with complex models or large datasets. & Fast, near real-time responses due to efficient language model inference. \\
\hline
\makecell{Adaptability} & Adaptable based on human expertise, but may require retraining or knowledge sharing. & Limited adaptability, requires retraining on new data or scenarios. & High adaptability to new scenarios and anomalies through prompt engineering and fine-tuning. \\
\hline
\makecell{Decision-Making Process} & Based on human expertise and judgement, may be subjective or inconsistent. & Black-box decision making, difficult to interpret or explain. & Explainable and transparent, with the ability to provide reasoning behind decisions. \\
\hline
\makecell{Error Rate} & Subject to human errors and biases, may vary based on operator expertise at each iteration. & Depends on model architecture and training data quality, may have FPs/FNs. & Low error rate due to robust language understanding and context-awareness. \\
\hline
\makecell{User Experience} & Requires interaction with human operators, may have communication challenges. & Often requires technical expertise to interpret results and adjust model parameters. & Intuitive and user-friendly natural language interface, requires minimal technical expertise. \\
\hline
\makecell{Learning Capabilities} & Relies on human learning and knowledge sharing, may be slower to adapt. & Requires retraining on new data, may have limited incremental learning capabilities. & Continual learning through interaction and fine-tuning, can rapidly adapt to new scenarios. \\
\hline
\makecell{Customization} & Customization depends on operator expertise and availability. & Can be customized by adjusting model architecture and hyperparameters, but may require significant effort. & Highly customizable through prompt engineering and fine-tuning for specific use cases. \\
\hline
\makecell{Data Handling} & Relies on human ability to interpret and analyze data, may be limited by data complexity. & Requires structured input data, may struggle with complex or unstructured datasets. & Can process and analyze unstructured text data, making it suitable for GOOSE/SV datasets. \\
\hline
\makecell{Scalability} & Limited scalability due to reliance on human operators, may become hindered. & Scalability depends on model complexity and computational resources, may require distributed training. & Highly scalable due to efficient inference and ability to handle large datasets. \\
\hline
\makecell{New Attack/Error} & Relies on human ability to identify and respond to new attacks or errors & May struggle to detect new attacks or errors not seen during training, requires retraining & Can quickly adapt to new attacks or errors through prompt engineering and fine-tuning. \\
\hline
\makecell{Complexity of Cases} & Complexity depends on operator expertise and experience, may require escalation for complex cases. & Limited complexity handling, may struggle with edge cases or complex anomalies. & Can handle complex cases through advanced language understanding and reasoning capabilities. \\
\hline
\end{tabular}
\end{table*}
The LLM-based ToD approach demonstrates several key advantages over its counterparts. It enables intuitive and user-friendly natural language interactions, reducing the need for extensive operator training and facilitating rapid adaptation to new scenarios through continual learning and fine-tuning. This adaptability allows the system to handle complex cases and respond effectively to new attacks. Moreover, the proposed framework can process and analyze unstructured text data, making it suitable for handling GOOSE/SV datasets, which may be complex. The system's scalability and ability to infer and adapt efficiently to new attacks or errors through prompt engineering ensure its robustness in detecting anomalies, overshadowing the capabilities of HITL and ML models~\cite{li2024personal}.
\section{Results and Discussion} \label{Results}
In this section, an extensive comparative analysis is conducted to assess the performance and efficacy of the proposed LLM-based ToD framework in relation to the HITL process across state-of-the-art LLMs (e.g., ChatGPT 4.0~\cite{chatgpt4.0}, Anthropic Claude Pro~\cite{anthropic}, Microsoft Copilot AI~\cite{Copilot}, and Google Bard/PaLM 2~\cite{googlebard}). This in-depth evaluation will shed light on the potential advantages and limitations of the proposed approach in contrast to the HITL process, offering valuable insights into the scalability and adaptability of these techniques in handling complex scenarios. By examining the performance metrics and benchmarking the results across these LLMs, this section aims to provide an understanding of the strengths and weaknesses of the proposed framework considering LLMs. The following parts show the results of the proposed AD, with their evaluation metrics compared with the LLM-based HITL method. Table~\ref{EvalMetric} shows the different descriptions and definitions for evaluation metrics to make a comparison between the proposed framework and the HITL process. 
\begin{table*}[]
\centering
\scriptsize
\caption{The performance evaluation metrics' definitions and representations.}
\label{EvalMetric}
\begin{tabular}{|l|l|l|}
\hline
\textbf{Metric}       & \textbf{Description} & \textbf{Formulation} \\ \hline
TP           &   True anomalies detected.          &     \cellcolor[HTML]{656565}        \\ \hline

TN           &   Normal communications correctly classified.         &     \cellcolor[HTML]{656565}        \\ \hline

FP           &    False anomalies detected.        &    \cellcolor[HTML]{656565}         \\ \hline

FN           &    True anomalies missed.         &    \cellcolor[HTML]{656565}         \\ \hline

TPR          &    The proportion of true anomalies that were correctly identified.         &   
 $\frac{TP}{TP+FN}$
\\ \hline

FPR          &   The rate of normal communications that were mistakenly identified as anomalies.          &    $\frac{FP}{FP+TN}$         \\ \hline

FNR          &     The rate of true anomalies that the system failed to detect.        &    $\frac{FN}{FN+TP}$         \\ \hline

Precision    &    The rate of detected anomalies that are indeed true anomalies.         &    $\frac{TP}{TP+FP}$         \\ \hline

Accuracy     &    The rate of correctly identified normal data and anomalies.         &     $\frac{TP+TN}{TP+TN+FP+FN}$        \\ \hline

F1-Score     &     It provides a balance between Precision and TPR.         &     $2 \times \frac{Precision \times TPR}{Precision + TPR}$        \\ \hline

Markedness   &   It measures the consistency of predictions.          &    $Precision + \frac{TN}{TN+FN}-1$         \\ \hline

Informedness &    It measures the probability of an informed decision.         &    $TPR+TNR-1$         \\ \hline

MCC          &   A balance measure of binary (normal or anomaly) classification quality.           &    $\frac{TP \times TN - FP \times FN}{\sqrt{(TP + FP) \times (TP + FN) \times (TN + FP) \times (TN + FN)}}$         \\ \hline
\end{tabular}
\end{table*}
According to this table, standard and advanced evaluation metrics are defined and formulated. It should be noted that the advanced metrics (i.e., Markedness, Informedness, Matthews Correlation Coefficient (MCC)) are employed in this paper to show the consistency, the decision-making process, and the quality of the classifications that are between $-1$ and $1$. These metrics can be useful in the AD process to check the capability and applicability of different LLMs according to the traditional and proposed methods. According to the application of AD in GOOSE/SV datasets, Markedness serves as a valuable metric for assessing the model's proficiency in mitigating both FPs (e.g., false alarms) and FNs (e.g., missed detection). A Markedness value approaching the upper end of its range signifies a highly dependable AD framework that effectively minimizes erroneous alerts. This characteristic is of paramount importance in ensuring the stability and optimal performance of substations by reducing unnecessary disruptions. Informedness indicates how well the model can detect variations in dataset patterns that signify anomalies. In the case of AD with LLMs, where the occurrence of actual anomalies may be rare, MCC is particularly useful. It ensures that the model’s performance is not overly influenced by the larger class size, providing a true indication of the model's effectiveness~\cite{de2022general}.

The aim of this section is to present the outcomes derived from the evaluation of metrics across various LLMs, categorized according to their respective training levels. The proposed LLM-based ToD framework has been implemented to show the results for evaluation metrics. Also, a comparative analysis of improvements of LLMs based on different training levels is considered for better understanding. Regarding the HITL process, three LLMs, including ChatGPT 4.0, Anthropic Claude Pro, and Google Bard/PaLM 2 (currently Gemini Advanced), have been implemented based on the IDS and human recommendations according to training levels. Then, the proposed framework was tested on two LLMs (Anthropic Claude Pro and Microsoft Copilot AI). The reason for other LLMs (e.g., ChatGPT 4.0, Gemini Advanced) to have not been considered for the proposed framework is that there were internal issues with these LLMs. According to the proposed framework, the interactions between humans and the LLM are based on prompt and response, and that was the target of using an image-based proposed framework to facilitate the process and require less effort. However, ChatGPT 4.0 and Gemini Advanced had issues with continuous errors in analyzing the images and encountered stopped analyses. Also, Gemini Advanced could not analyze the images according to its features (it is a text-based LLM). The next part presents the results and compares between the HITL process and the proposed model based on performance evaluation metrics.
\subsection{Case Studies: SV and GOOSE Anomaly Detection} \label{ResultsCharts}
\begin{figure}[!h]
\centerline{\includegraphics[width=0.9\columnwidth]{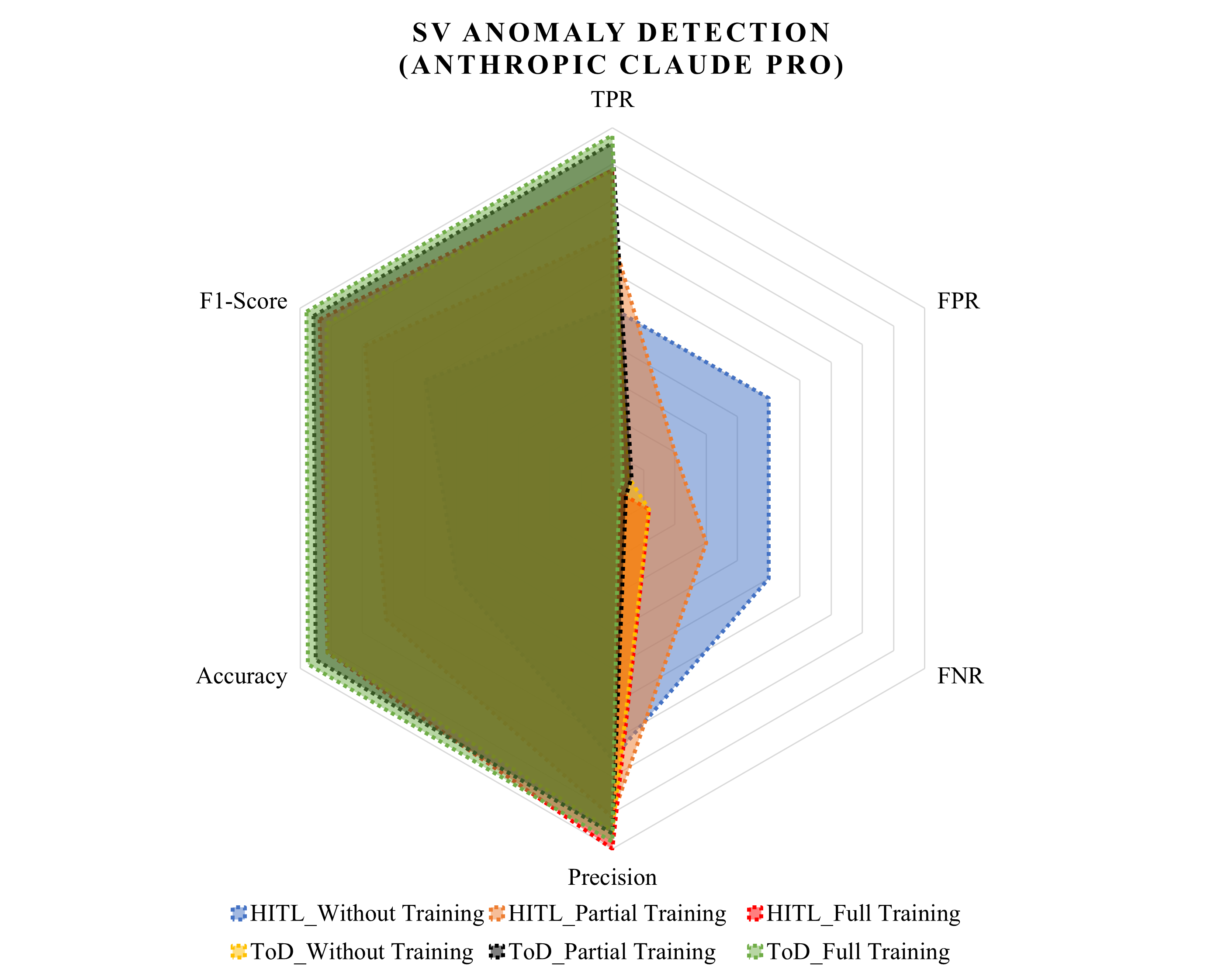}}
\caption{A comparison of HITL process and the proposed methodology in AD for SV datasets in the same LLM (i.e, Anthropic Claude Pro).}
\label{SV_HITL&ToD}
\end{figure}

\begin{figure}[!h]
\centerline{\includegraphics[width=0.9\columnwidth]{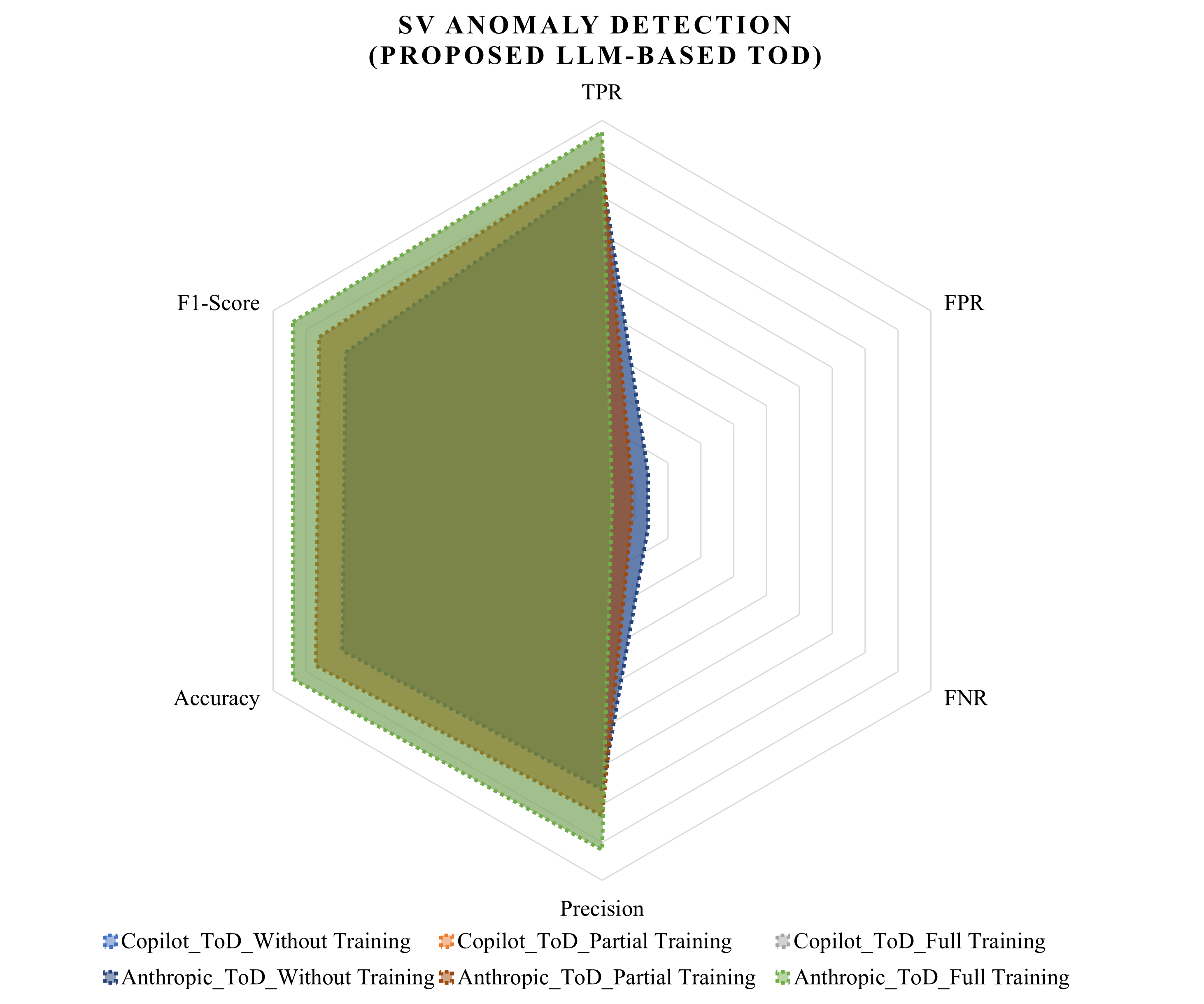}}
\caption{A comparative LLM-based ToD framework for SV dataset.}
\label{SV_OnlyToDs}
\end{figure}

\begin{figure}[!h]
\centerline{\includegraphics[width=0.9\columnwidth]{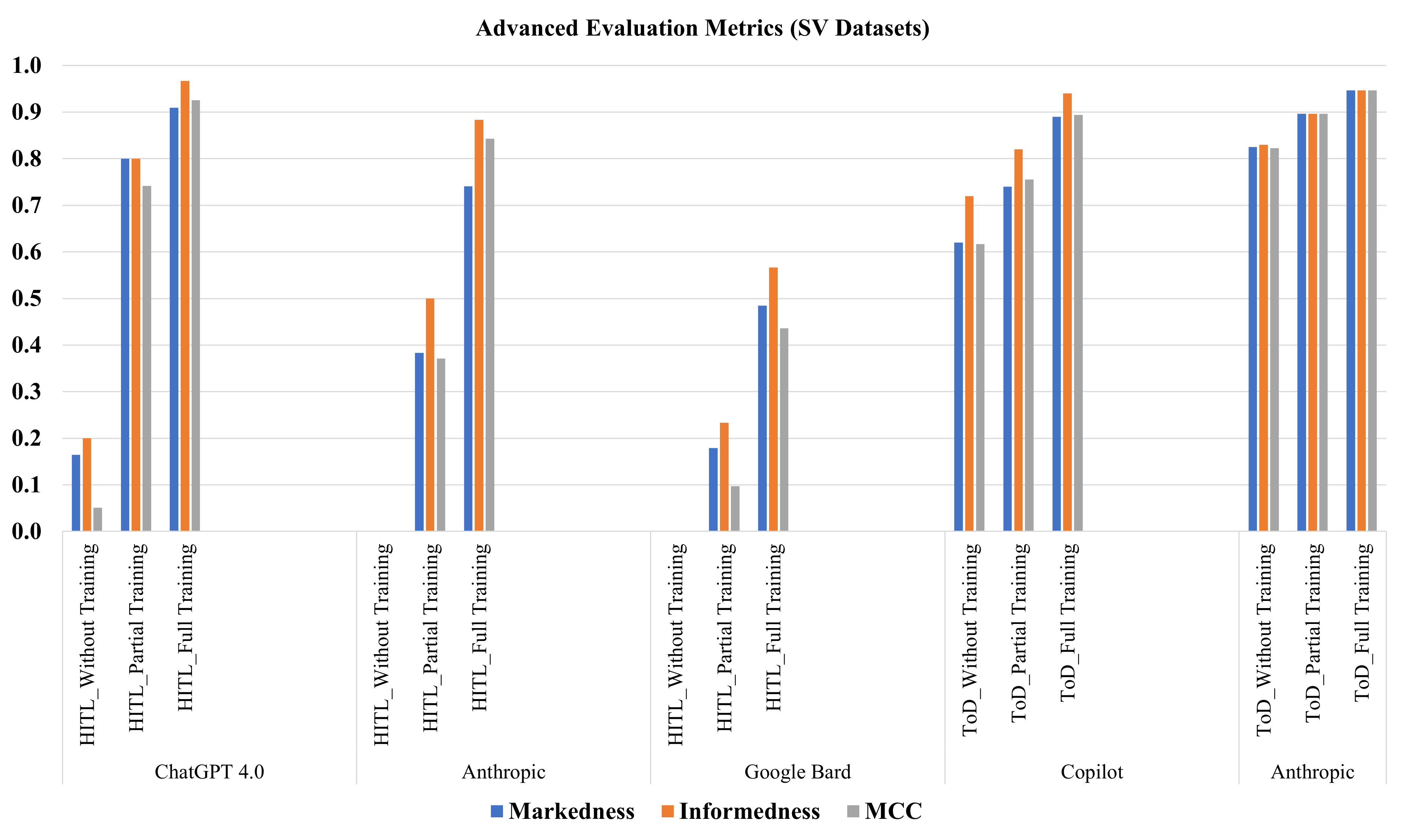}}
\caption{A comparison of advanced evaluation metrics for SV datasets considering the HITL process and the proposed framework using LLMs.}
\label{AdvancedMetricSV}
\end{figure}

This section compares different IDSs based on the HITL method and the proposed LLM-based ToD framework according to comprehensive evaluation metrics in various LLMs. Two datasets are considered for SV and GOOSE multicast messages. As shown in Fig.~\ref{SV_HITL&ToD}, Anthropic LLM (green color) shows superior performance according to the proposed framework, compared with the LLM-based HITL model based on the metrics. 
The \textit{ToD\_Full Training} mode has the best performance in terms of the highest TPR, Precision, Accuracy, and F1-Score in the SV dataset. In addition, the lowest percentage of FPR and FNR roughly happened in the proposed framework implemented in Anthropic LLM. Also, the \textit{HITL\_Without Training} mode has the lowest efficiency and performance, with a relatively low accuracy of $50\%$. According to Table~\ref{results}, it should be noted that ChatGPT 4.0 showed approximately good results in the full training level for the SV dataset in the HITL process and has very close performance to that of the proposed model implemented in the Anthropic. However, the proposed framework generally performs better considering all training levels. This paper considers advanced evaluation metrics, in which the Anthropic and Google Bard LLMs have values of $0$ for these metrics (as shown in Table~\ref{results}) in the HITL process. That means the model's performance in predicting positives and negatives is moderately no better than random guessing, and these models have a high rate of FPs and FNs, affecting the correct predictions. Also, the model's ability to make informed decisions based on the available SV dataset is not better than flipping a coin. Thus, these values show that these models cannot distinguish between normal and abnormal instances. Hence, these models can be considered unreliable as there is no good correlation between them and the identification of correct predictions. Also, this happened in the MCC metric of the GOOSE dataset in the HITL process of Google Bard LLM. Fig.~\ref{SV_OnlyToDs} compares two implemented ToD frameworks in Anthropic and Copilot LLMs. 
As can be seen, Anthropic outperforms Copilot in almost all metrics for SV datasets. Also, the full training mode of this LLM presents the best performance, with Accuracy, Precision, and F1-Score near $98\%$, which is satisfactory. Finally, the results of advanced metrics depicted in Fig.~\ref{AdvancedMetricSV} show that the proposed framework implemented by Anthropic has the highest performance on average and values close to $1$, which makes this model efficient, scalable, adaptable, and reliable. 
Many research studies ignored advanced metrics in their results that jeopardized their model's efficiency and reliability in the AD process.

\begin{figure}[!h]
\centerline{\includegraphics[width=0.9\columnwidth]{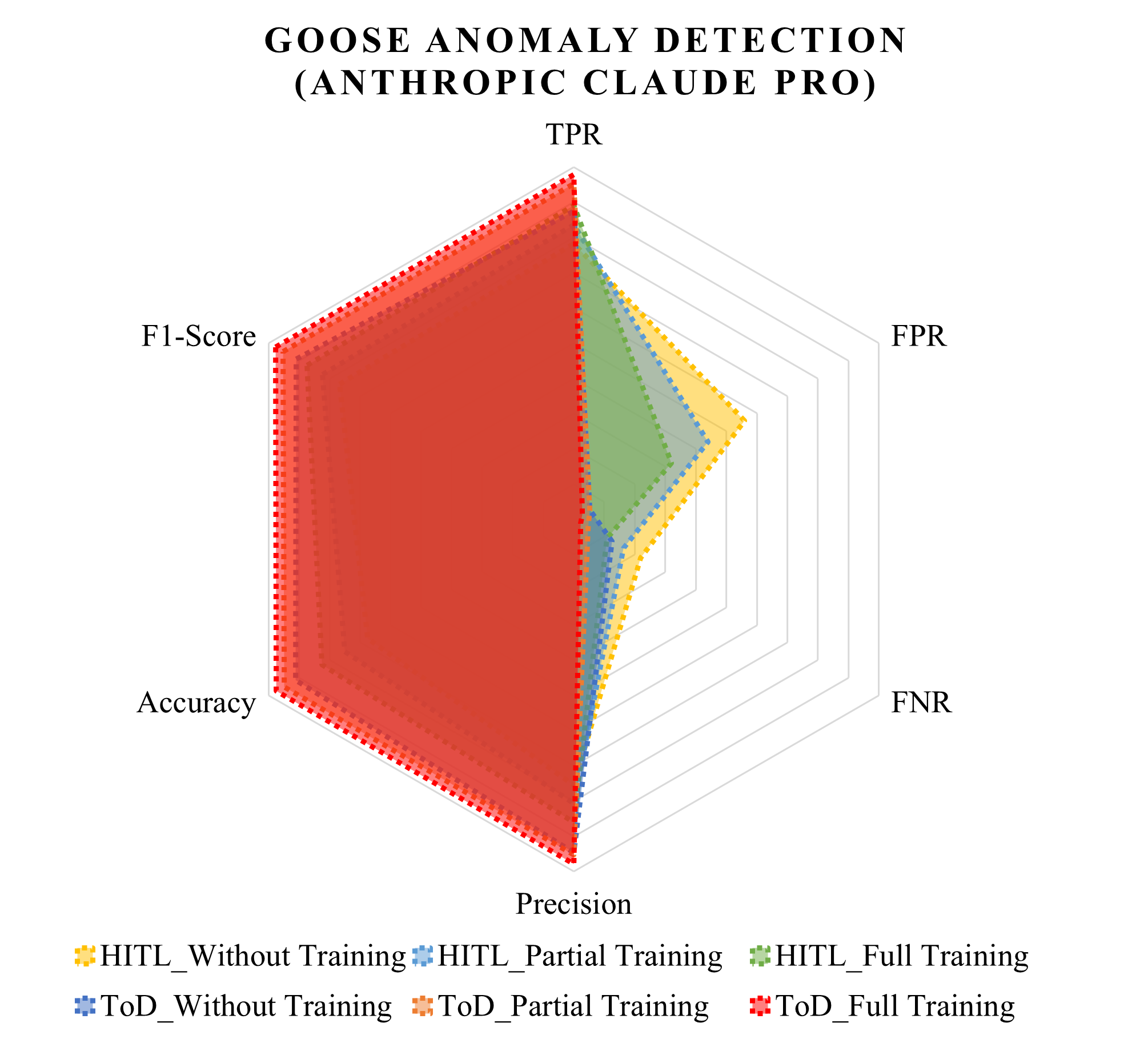}}
\caption{A comparison of HITL process and the proposed methodology in AD for GOOSE datasets in same LLM (i.e, Anthropic Claude Pro).}
\label{GOOSE_HITL&ToD}
\end{figure}

\begin{figure}[!h]
\centerline{\includegraphics[width=0.9\columnwidth]{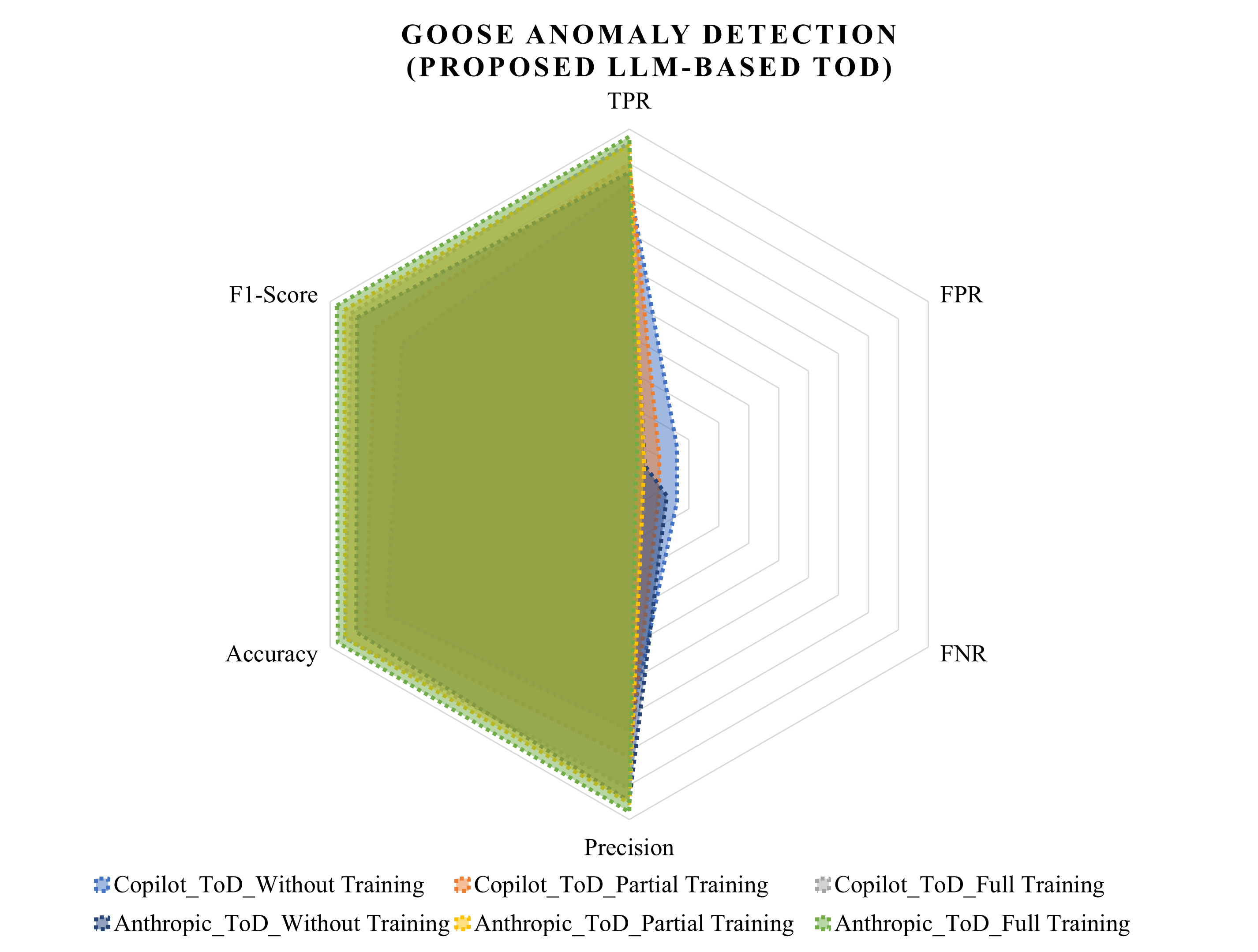}}
\caption{A comparative LLM-based ToD framework for GOOSE dataset.}
\label{GOOSE_OnlyToDs}
\end{figure}

\begin{figure}[!h]
\centerline{\includegraphics[width=0.9\columnwidth]{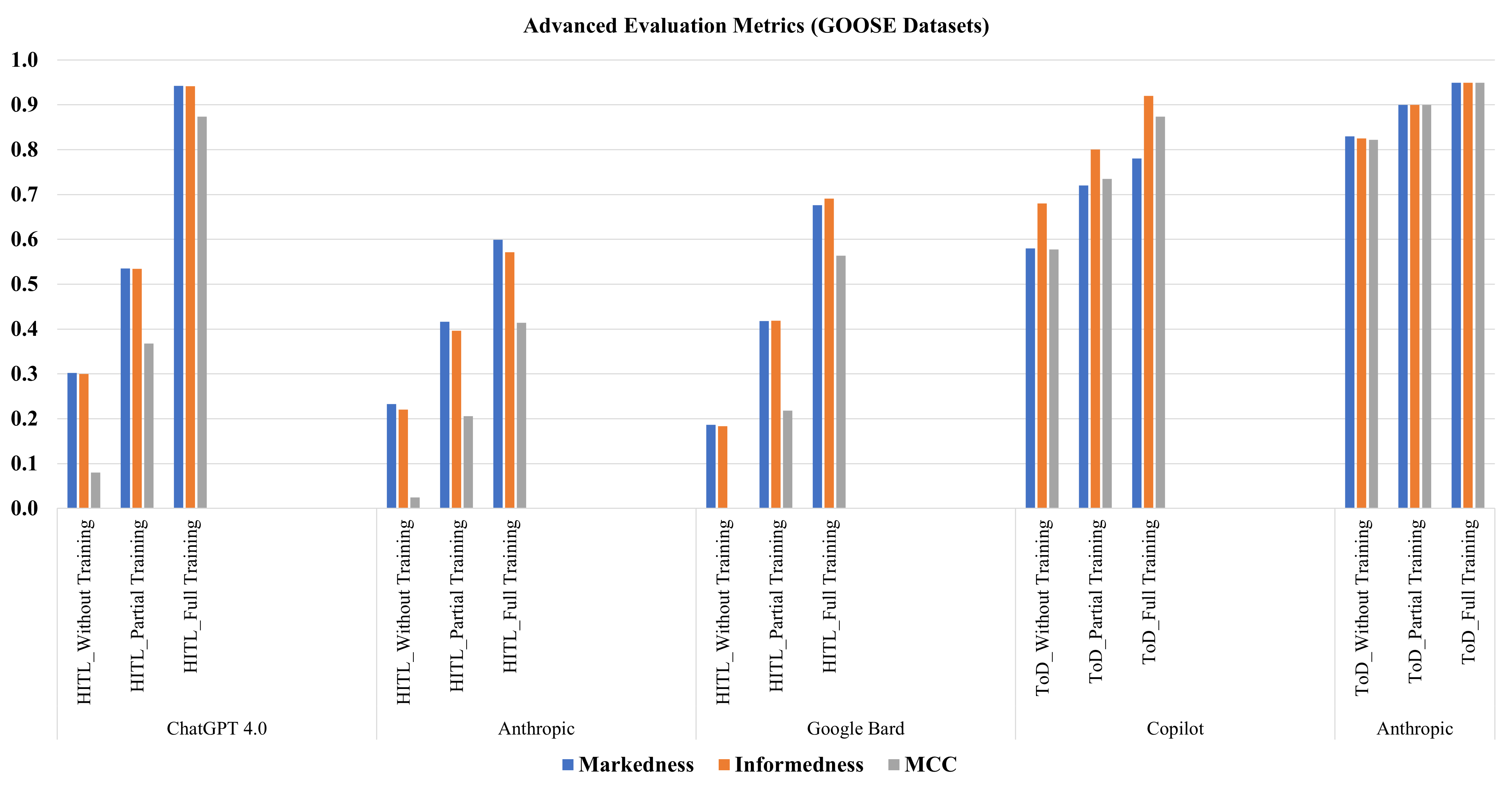}}
\caption{A comparison of advanced evaluation metrics for GOOSE datasets considering the HITL process and the proposed framework using LLMs.}
\label{AdvancedMetricGOOSE}
\end{figure}

\begin{figure}[!h]
\centerline{\includegraphics[width=0.8\columnwidth]{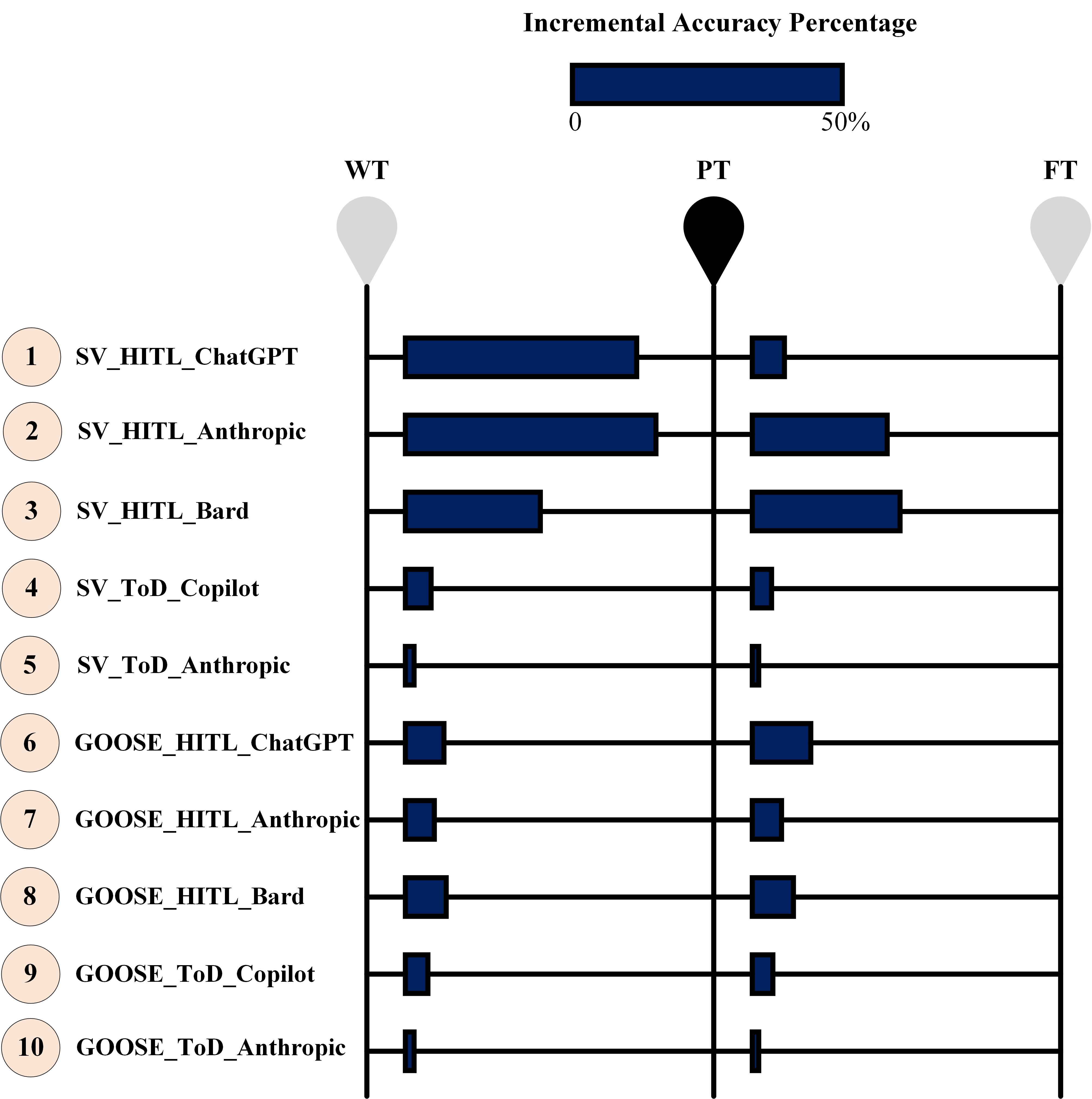}}
\caption{An incremental accuracy percentage of different models for training levels (i.e., WT: without training, PT: partial training, FT: full training).}
\label{AccuracyDiff}
\end{figure}

Employing GOOSE datasets, a comparison of the HITL process and the proposed ToD model is presented in Fig.~\ref{GOOSE_HITL&ToD} and the bottom part of Table~\ref{results}. According to the figure, \textit{ToD\_Full Training} demonstrated the best performance in almost all metrics in general. However, ChatGPT 4.0 has good results, based only on the full training of the HITL process (as bolded in the table), which is very close to the proposed framework by Anthropic LLM. As mentioned before, ChatGPT 4.0 could not handle the proposed framework based on the human interactions and was very slow to handle this task. On the other hand, Anthropic was very user-friendly and manageable to understand and interpret different prompts. 
Based on Fig.~\ref{GOOSE_OnlyToDs}, the green part (i.e., \textit{Anthropic\_ToD\_Full Training} mode) outperforms other proposed models in different training levels. Although, the proposed framework implemented in Copilot has good results in comparison with HITL models, the Anthropic implementation, however, showed outstanding results even in the without-training part, which is better than the partial training of the proposed model in Copilot in terms of Accuracy, Precision, and F1-Score metrics.
Lastly, an advanced metric analysis carried out according to Fig.~\ref{AdvancedMetricGOOSE} shows that the Anthropic model based on the framework is superior in all metrics. These analyses demonstrate the reliability and robustness of the proposed model in the AD concept for GOOSE datasets. 
Another feature is illustrated in Fig.~\ref{AccuracyDiff} that can be useful in comparing different models deployed in different training levels and LLMs. 
This diagram depicts all cases, including the HITL and proposed models along with their training levels, for GOOSE and SV datasets, based on the Accuracy metric. The purpose of this figure is to check the incremental percentage at each level. As can be seen, Anthropic has a very small increase according to the proposed framework, which means this model can make a good correlation between different training levels. Also, it can detect anomalies at a very good rate even without any training process, as there is not much difference between assorted levels for both SV and GOOSE datasets. The highest incremental percentage is assigned to the HITL process of Anthropic, which can be more adaptable to learn new recommendations, and next is ChatGPT 4.0. Further, this figure visualizes the accuracy differences in models to check their incremental percentage in comparison with the previous training level.

To recap, the LLM-based ToD framework executed in Anthropic Claude Pro showed the most efficient performance, with high reliability, scalability, and adaptability compared with HITL models and the proposed model carried out by Copilot LLM. ChatGPT 4.0 and Gemini Advanced (formerly Google Bard) did not perform well in implementing the proposed framework because of internal errors, and they stopped analyzing or rejected image uploads, making it hard to interact. According to datasets from the testbed and the proposed framework, it can be deduced that Anthropic Claude Pro currently performs best in the AD process of multicast messages. Also, the LLM-based ToD framework needs minor effort in comparison with ML algorithms because it does not require re-training in cases of new anomalies as LLM-based models have the feature of language processing that make the interpretation and detection more feasible.
\begin{table*}[h]
\centering
\fontsize{6pt}{8pt}\selectfont
\caption{A comparison of AD results (``without,'' ``partial,'' and ``full'' terms show the levels of training process).}
\label{results}
\begin{tabular}{|c|ccccccccccccccc|}
\hline
\cellcolor[HTML]{656565}  & \multicolumn{15}{c|}{\cellcolor[HTML]{FFFFC7}{\color[HTML]{000000} \textbf{SV}}}                                        \\ \hline
\makecell{\textbf{LLMs}}      
& \multicolumn{3}{c|}{HITL (ChatGPT 4.0)}                                                        & \multicolumn{3}{c|}{HITL (Anthropic's Claude Pro)}                                                        & \multicolumn{3}{c|}{HITL (Google Bard/PaLM 2)}     &     \multicolumn{3}{c|}{ToD (Microsoft Copilot AI)}  & \multicolumn{3}{c|}{ToD (Anthropic's Claude Pro)}         \\ \hline
      \textbf{\makecell{Standard Metrics}}    & \multicolumn{1}{c|}{without} & \multicolumn{1}{c|}{partial} & \multicolumn{1}{c|}{full} & \multicolumn{1}{c|}{without} & \multicolumn{1}{c|}{partial} & \multicolumn{1}{c|}{full} & \multicolumn{1}{c|}{without} & \multicolumn{1}{c|}{partial} & \multicolumn{1}{c|}{full} & \multicolumn{1}{c|}{without} & \multicolumn{1}{c|}{partial} & \multicolumn{1}{c|}{full} & \multicolumn{1}{c|}{without} & \multicolumn{1}{c|}{partial} &  full \\ \hline
\textit{TPR}      & \multicolumn{1}{c|}{70\%}        & \multicolumn{1}{c|}{95\%}        & \multicolumn{1}{c|}{96.67\%}     & \multicolumn{1}{c|}{50\%}        & \multicolumn{1}{c|}{70\%}        & \multicolumn{1}{c|}{88.3\%}     & \multicolumn{1}{c|}{50\%}        & \multicolumn{1}{c|}{63.3\%}   & \multicolumn{1}{c|}{81.6\%} & \multicolumn{1}{c|}{86\%} & \multicolumn{1}{c|}{91\%} & \multicolumn{1}{c|}{97\%} & \multicolumn{1}{c|}{\textbf{88.37\%}} & \multicolumn{1}{c|}{\textbf{95.74\%}}     & \textbf{98\%}  \\ \hline
\textit{FPR}      & \multicolumn{1}{c|}{50\%}        & \multicolumn{1}{c|}{15\%}        & \multicolumn{1}{c|}{\textbf{0\%}}     & \multicolumn{1}{c|}{50\%}        & \multicolumn{1}{c|}{20\%}        & \multicolumn{1}{c|}{0\%}     & \multicolumn{1}{c|}{50\%}        & \multicolumn{1}{c|}{40\%}   & \multicolumn{1}{c|}{25\%} & \multicolumn{1}{c|}{14\%} & \multicolumn{1}{c|}{9\%} & \multicolumn{1}{c|}{3\%} & \multicolumn{1}{c|}{\textbf{5.41\%}} & \multicolumn{1}{c|}{\textbf{6.06\%}}     &    3.33\%  \\ \hline
\textit{FNR}       & \multicolumn{1}{c|}{30\%}        & \multicolumn{1}{c|}{5\%}        & \multicolumn{1}{c|}{3.33\%}     & \multicolumn{1}{c|}{50\%}        & \multicolumn{1}{c|}{30\%}        & \multicolumn{1}{c|}{11.67\%}     & \multicolumn{1}{c|}{50\%}        & \multicolumn{1}{c|}{36.6\%}   & \multicolumn{1}{c|}{18.34\%} & \multicolumn{1}{c|}{14\%} & \multicolumn{1}{c|}{9\%} & \multicolumn{1}{c|}{3\%} & \multicolumn{1}{c|}{\textbf{11.63\%}} & \multicolumn{1}{c|}{\textbf{4.26\%}}     &   \textbf{2\%}   \\ \hline
\textit{Precision} & \multicolumn{1}{c|}{80.77\%}        & \multicolumn{1}{c|}{95\%}        & \multicolumn{1}{c|}{\textbf{100\%}}     & \multicolumn{1}{c|}{75\%}        & \multicolumn{1}{c|}{91.3\%}        & \multicolumn{1}{c|}{100\%}     & \multicolumn{1}{c|}{75\%}        & \multicolumn{1}{c|}{82.6\%}   & \multicolumn{1}{c|}{91.7\%} & \multicolumn{1}{c|}{76\%} & \multicolumn{1}{c|}{83\%} & \multicolumn{1}{c|}{92\%} & \multicolumn{1}{c|}{\textbf{95\%}} & \multicolumn{1}{c|}{\textbf{95.74\%}}     &   98\%   \\ \hline

\textit{Accuracy}      & \multicolumn{1}{c|}{65\%}        & \multicolumn{1}{c|}{92.5\%}        & \multicolumn{1}{c|}{97.5\%}     & \multicolumn{1}{c|}{50\%}        & \multicolumn{1}{c|}{72.5\%}        & \multicolumn{1}{c|}{91.25\%}     & \multicolumn{1}{c|}{50\%}        & \multicolumn{1}{c|}{62.5\%}   & \multicolumn{1}{c|}{80\%} & \multicolumn{1}{c|}{79\%} & \multicolumn{1}{c|}{87\%} & \multicolumn{1}{c|}{94\%} & \multicolumn{1}{c|}{\textbf{91.25\%}} & \multicolumn{1}{c|}{\textbf{95\%}}     & \textbf{97.5\% } \\ \hline

\textit{F1-Score}  & \multicolumn{1}{c|}{75\%}        & \multicolumn{1}{c|}{95\%}        & \multicolumn{1}{c|}{98.3\%}     & \multicolumn{1}{c|}{60\%}        & \multicolumn{1}{c|}{79.2\%}        & \multicolumn{1}{c|}{93.8\%}     & \multicolumn{1}{c|}{60\%}        & \multicolumn{1}{c|}{71.7\%}   & \multicolumn{1}{c|}{85.9\%} & \multicolumn{1}{c|}{78\%} & \multicolumn{1}{c|}{86\%} & \multicolumn{1}{c|}{94\%} & \multicolumn{1}{c|}{\textbf{91.57\%}} & \multicolumn{1}{c|}{\textbf{95.74\%}}     &   \textbf{98.4\%}  \\ \hline

\textbf{Advanced Metrics}  & \multicolumn{1}{c|}{\cellcolor{gray!30}}        & \multicolumn{1}{c|}{\cellcolor{gray!30}}        & \multicolumn{1}{c|}{\cellcolor{gray!30}}     & \multicolumn{1}{c|}{\cellcolor{gray!30}}        & \multicolumn{1}{c|}{\cellcolor{gray!30}}        & \multicolumn{1}{c|}{\cellcolor{gray!30}}     & \multicolumn{1}{c|}{\cellcolor{gray!30}}        & \multicolumn{1}{c|}{\cellcolor{gray!30}}   & \multicolumn{1}{c|}{\cellcolor{gray!30}} & \multicolumn{1}{c|}{\cellcolor{gray!30}} & \multicolumn{1}{c|}{\cellcolor{gray!30}} & \multicolumn{1}{c|}{\cellcolor{gray!30}} & \multicolumn{1}{c|}{\cellcolor{gray!30}} & \multicolumn{1}{c|}{\cellcolor{gray!30}}     &   \cellcolor{gray!30}  \\ \hline

\textit{Markedness}  & \multicolumn{1}{c|}{0.1648}        & \multicolumn{1}{c|}{0.8}        & \multicolumn{1}{c|}{0.9091}     & \multicolumn{1}{c|}{\textbf{0}}        & \multicolumn{1}{c|}{0.3836}        & \multicolumn{1}{c|}{0.7407}     & \multicolumn{1}{c|}{\textbf{0}}        & \multicolumn{1}{c|}{0.1789}   & \multicolumn{1}{c|}{0.4843} & \multicolumn{1}{c|}{0.62} & \multicolumn{1}{c|}{0.74} & \multicolumn{1}{c|}{0.89} & \multicolumn{1}{c|}{\textbf{0.825}} & \multicolumn{1}{c|}{\textbf{0.8968}}     &   \textbf{0.9467} \\ \hline

\textit{Informedness}  & \multicolumn{1}{c|}{0.2}        & \multicolumn{1}{c|}{0.8}        & \multicolumn{1}{c|}{\textbf{0.9667}}     & \multicolumn{1}{c|}{\textbf{0}}        & \multicolumn{1}{c|}{0.5}        & \multicolumn{1}{c|}{0.8833}     & \multicolumn{1}{c|}{\textbf{0}}        & \multicolumn{1}{c|}{0.2333}   & \multicolumn{1}{c|}{0.5667} & \multicolumn{1}{c|}{0.72} & \multicolumn{1}{c|}{0.82} & \multicolumn{1}{c|}{0.94} & \multicolumn{1}{c|}{\textbf{0.8296}} & \multicolumn{1}{c|}{\textbf{0.8968}}     &   0.9467 \\ \hline

\textit{MCC}  & \multicolumn{1}{c|}{0.0513}        & \multicolumn{1}{c|}{0.7416}        & \multicolumn{1}{c|}{0.9258}     & \multicolumn{1}{c|}{\textbf{0}}        & \multicolumn{1}{c|}{0.3713}        & \multicolumn{1}{c|}{0.8432}     & \multicolumn{1}{c|}{\textbf{0}}        & \multicolumn{1}{c|}{0.0976}   & \multicolumn{1}{c|}{0.4364} & \multicolumn{1}{c|}{0.6164} & \multicolumn{1}{c|}{0.7554} & \multicolumn{1}{c|}{0.8944} & \multicolumn{1}{c|}{\textbf{0.823}} & \multicolumn{1}{c|}{\textbf{0.8966}}     &  \textbf{0.9466}  \\ \hline
\cellcolor[HTML]{656565}  & \multicolumn{15}{c|}{\cellcolor[HTML]{FFFFC7}{\color[HTML]{000000} \textbf{GOOSE}}}                                        \\ \hline
\makecell{\textbf{LLMs}}      
& \multicolumn{3}{c|}{HITL (ChatGPT 4.0)}                                                        & \multicolumn{3}{c|}{HITL (Anthropic's Claude Pro)}                                                        & \multicolumn{3}{c|}{HITL (Google Bard/PaLM 2)}     &     \multicolumn{3}{c|}{ToD (Microsoft Copilot AI)}  & \multicolumn{3}{c|}{ToD (Anthropic's Claude Pro)}                          \\ \hline
      \textbf{\makecell{Standard Metrics}}    & \multicolumn{1}{c|}{without} & \multicolumn{1}{c|}{partial} & \multicolumn{1}{c|}{full} & \multicolumn{1}{c|}{without} & \multicolumn{1}{c|}{partial} & \multicolumn{1}{c|}{full} & \multicolumn{1}{c|}{without} & \multicolumn{1}{c|}{partial} & \multicolumn{1}{c|}{full} & \multicolumn{1}{c|}{without} & \multicolumn{1}{c|}{partial} & \multicolumn{1}{c|}{full} & \multicolumn{1}{c|}{without} & \multicolumn{1}{c|}{partial} &  full \\ \hline
\textit{TPR}       & \multicolumn{1}{c|}{78.18\%}        & \multicolumn{1}{c|}{85.45\%}        & \multicolumn{1}{c|}{\textbf{98.18\%}}     & \multicolumn{1}{c|}{78.18\%}        & \multicolumn{1}{c|}{83.64\%}        & \multicolumn{1}{c|}{89.09\%}     & \multicolumn{1}{c|}{74.5\%}        & \multicolumn{1}{c|}{81.8\%}   & \multicolumn{1}{c|}{89.1\%} & \multicolumn{1}{c|}{84\%} & \multicolumn{1}{c|}{90\%} & \multicolumn{1}{c|}{96\%} & \multicolumn{1}{c|}{\textbf{87.5\%}} & \multicolumn{1}{c|}{\textbf{95.24\%}}     & 97.78\%  \\ \hline
\textit{FPR}       & \multicolumn{1}{c|}{48\%}        & \multicolumn{1}{c|}{32\%}        & \multicolumn{1}{c|}{4\%}     & \multicolumn{1}{c|}{56\%}        & \multicolumn{1}{c|}{44\%}        & \multicolumn{1}{c|}{32\%}     & \multicolumn{1}{c|}{56\%}        & \multicolumn{1}{c|}{40\%}   & \multicolumn{1}{c|}{20\%} & \multicolumn{1}{c|}{16\%} & \multicolumn{1}{c|}{10\%} & \multicolumn{1}{c|}{4\%} & \multicolumn{1}{c|}{\textbf{5.26\%}} & \multicolumn{1}{c|}{\textbf{5\%}}     &    \textbf{2.86\%}  \\ \hline
\textit{FNR}       & \multicolumn{1}{c|}{21.82\%}        & \multicolumn{1}{c|}{14.55\%}        & \multicolumn{1}{c|}{\textbf{1.82\%}}     & \multicolumn{1}{c|}{21.82\%}        & \multicolumn{1}{c|}{16.36\%}        & \multicolumn{1}{c|}{10.91\%}     & \multicolumn{1}{c|}{25.5\%}        & \multicolumn{1}{c|}{18.18\%}   & \multicolumn{1}{c|}{10.9\%} & \multicolumn{1}{c|}{16\%} & \multicolumn{1}{c|}{10\%} & \multicolumn{1}{c|}{4\%} & \multicolumn{1}{c|}{\textbf{12.5\%}} & \multicolumn{1}{c|}{\textbf{4.76\%}}     &   2.22\%   \\ \hline
\textit{Precision} & \multicolumn{1}{c|}{78.18\%}        & \multicolumn{1}{c|}{85.45\%}        & \multicolumn{1}{c|}{\textbf{98.18\%}}     & \multicolumn{1}{c|}{75.43\%}        & \multicolumn{1}{c|}{80.7\%}        & \multicolumn{1}{c|}{85.96\%}     & \multicolumn{1}{c|}{74.5\%}        & \multicolumn{1}{c|}{81.8\%}   & \multicolumn{1}{c|}{90.7\%} & \multicolumn{1}{c|}{74\%} & \multicolumn{1}{c|}{82\%} & \multicolumn{1}{c|}{91\%} & \multicolumn{1}{c|}{\textbf{94.59\%}} & \multicolumn{1}{c|}{\textbf{95.24\%}}     &   97.78\%   \\ \hline

\textit{Accuracy}      & \multicolumn{1}{c|}{70\%}        & \multicolumn{1}{c|}{80\%}        & \multicolumn{1}{c|}{97.5\%}     & \multicolumn{1}{c|}{67.5\%}        & \multicolumn{1}{c|}{75\%}        & \multicolumn{1}{c|}{82.5\%}     & \multicolumn{1}{c|}{65\%}        & \multicolumn{1}{c|}{75\%}   & \multicolumn{1}{c|}{86.25\%} & \multicolumn{1}{c|}{81\%} & \multicolumn{1}{c|}{88\%} & \multicolumn{1}{c|}{95\%} & \multicolumn{1}{c|}{\textbf{91.25\%}} & \multicolumn{1}{c|}{\textbf{95\%}}     & \textbf{97.5\%}  \\ \hline

\textit{F1-Score}  & \multicolumn{1}{c|}{78.18\%}        & \multicolumn{1}{c|}{85.45\%}        & \multicolumn{1}{c|}{\textbf{98.18\%}}     & \multicolumn{1}{c|}{76.78\%}        & \multicolumn{1}{c|}{82.3\%}        & \multicolumn{1}{c|}{87.5\%}     & \multicolumn{1}{c|}{74.5\%}        & \multicolumn{1}{c|}{81.8\%}   & \multicolumn{1}{c|}{90.7\%} & \multicolumn{1}{c|}{76\%} & \multicolumn{1}{c|}{85\%} & \multicolumn{1}{c|}{93\%} & \multicolumn{1}{c|}{\textbf{90.91\%}} & \multicolumn{1}{c|}{\textbf{95.24\%}}     &   97.78\%   \\ \hline

\textbf{Advanced Metrics}  & \multicolumn{1}{c|}{\cellcolor{gray!30}}        & \multicolumn{1}{c|}{\cellcolor{gray!30}}        & \multicolumn{1}{c|}{\cellcolor{gray!30}}     & \multicolumn{1}{c|}{\cellcolor{gray!30}}        & \multicolumn{1}{c|}{\cellcolor{gray!30}}        & \multicolumn{1}{c|}{\cellcolor{gray!30}}     & \multicolumn{1}{c|}{\cellcolor{gray!30}}        & \multicolumn{1}{c|}{\cellcolor{gray!30}}   & \multicolumn{1}{c|}{\cellcolor{gray!30}} & \multicolumn{1}{c|}{\cellcolor{gray!30}} & \multicolumn{1}{c|}{\cellcolor{gray!30}} & \multicolumn{1}{c|}{\cellcolor{gray!30}} & \multicolumn{1}{c|}{\cellcolor{gray!30}} & \multicolumn{1}{c|}{\cellcolor{gray!30}}     &   \cellcolor{gray!30}  \\ \hline

\textit{Markedness}  & \multicolumn{1}{c|}{0.302}        & \multicolumn{1}{c|}{0.535}        & \multicolumn{1}{c|}{0.942}     & \multicolumn{1}{c|}{0.233}        & \multicolumn{1}{c|}{0.416}        & \multicolumn{1}{c|}{0.599}     & \multicolumn{1}{c|}{0.186}        & \multicolumn{1}{c|}{0.418}   & \multicolumn{1}{c|}{0.676} & \multicolumn{1}{c|}{0.58} & \multicolumn{1}{c|}{0.72} & \multicolumn{1}{c|}{0.87} & \multicolumn{1}{c|}{\textbf{0.8296}} & \multicolumn{1}{c|}{\textbf{0.8998}}     &  \textbf{0.9492}  \\ \hline

\textit{Informedness}  & \multicolumn{1}{c|}{0.3}        & \multicolumn{1}{c|}{0.5345}        & \multicolumn{1}{c|}{0.9418}     & \multicolumn{1}{c|}{0.22}        & \multicolumn{1}{c|}{0.3964}        & \multicolumn{1}{c|}{0.5709}     & \multicolumn{1}{c|}{0.1836}        & \multicolumn{1}{c|}{0.4182}   & \multicolumn{1}{c|}{0.6909} & \multicolumn{1}{c|}{0.68} & \multicolumn{1}{c|}{0.8} & \multicolumn{1}{c|}{0.92} & \multicolumn{1}{c|}{\textbf{0.825}} & \multicolumn{1}{c|}{\textbf{0.8998}}     &  \textbf{0.9492}  \\ \hline

\textit{MCC}  & \multicolumn{1}{c|}{0.0798}        & \multicolumn{1}{c|}{0.3676}        & \multicolumn{1}{c|}{0.8737}     & \multicolumn{1}{c|}{0.0247}        & \multicolumn{1}{c|}{0.2054}        & \multicolumn{1}{c|}{0.4142}     & \multicolumn{1}{c|}{\textbf{0}}        & \multicolumn{1}{c|}{0.2182}   & \multicolumn{1}{c|}{0.5636} & \multicolumn{1}{c|}{0.5772} & \multicolumn{1}{c|}{0.7348} & \multicolumn{1}{c|}{0.8737} & \multicolumn{1}{c|}{\textbf{0.822}} & \multicolumn{1}{c|}{\textbf{0.8997}}     &  \textbf{0.9491}  \\ \hline
\end{tabular}
\end{table*}
\section{Conclusion} \label{Conclusion}
This paper proposes for the first time a novel LLM-based ToD framework that can detect anomalies in multicast messages in digital substations in a very efficient and reliable way. A comparison of the LLM-based proposed model with other HITL processes has been carried out to validate its scalability and adaptability mathematically and conceptually in parallel with HITL and ML methods. Also, advanced evaluation metrics are regarded in this paper to check the reliability and correlation ability of the LLM-based models, whereas previous research neglected these metrics in the smart grid domain. Anthropic Claude Pro demonstrates the best general results compared with other HITL processes or Copilot-based ToD frameworks. This innovative framework was implemented in the most famous LLMs, and the proposed methodology showed the best results in almost all metrics, even without any human recommendation at the training level.

In the future, the main goal will be to develop this framework by adding a self-learning block to collect additional recommendations and make a comprehensive dataset regarding other multicast messages, including MMS, Simple Network Time Protocol (SNTP), Precision Time Protocol (PTP), etc. Also, the quality of generated texts will be assessed by natural language processing to increase the acceptability of LLMs.

\bibliographystyle{IEEEtran}
\bibliography{IEEEabrv,ref}

\end{document}